\documentclass[journal,10pt]{IEEEtran}

\usepackage[T1]{fontenc}
\usepackage[utf8]{inputenc}
\usepackage{amsmath,amssymb}
\usepackage{graphicx,adjustbox}
\usepackage{booktabs,array,tabularx}
\usepackage{listings}
\usepackage{microtype}
\usepackage{balance}
\usepackage{xurl}
\usepackage[hidelinks]{hyperref}
\usepackage{tikz}
\usetikzlibrary{arrows.meta,positioning,fit}
\providecommand{\tightlist}{\setlength{\itemsep}{0pt}\setlength{\parskip}{0pt}}
\newcommand{\fitdisplay}[1]{\begin{equation*}\begin{adjustbox}{max width=\linewidth}$\displaystyle #1$\end{adjustbox}\end{equation*}}
\title{Verification Reward Model for Reinforcement Learning in Chip Design Verification}
\author{Shashank~Chaurasia%
\thanks{S. Chaurasia is an AI Researcher with MooresLab AI (e-mail: \url{shashank@mooreslab.ai}). Preprint, September 2026.}}
\hypersetup{pdftitle={Verification Reward Model for Reinforcement Learning in Chip Design Verification},pdfauthor={Shashank Chaurasia},pdfsubject={Position paper on reward modeling for LLM-generated chip verification artifacts}}
\begin{document}
\maketitle
\begin{abstract}
We propose a Verification Reward Model (VRM) framework for training language models to create and repair chip verification artifacts. The central object is a versioned verification contract that binds requirements, permissible stimulus, observation boundaries, reference behavior, evaluation budgets, and acceptance criteria. Compiler, simulator, formal, mutation, coverage, and expert-review evidence are converted into auditable records. Deterministic acceptance checks remain outside the learned model. A learned outcome model predicts expensive future evidence from the specification, the generated artifact, and explicitly masked partial evidence; a semantic critic identifies evidence-supported weaknesses; a deterministic reward composer translates the resulting quality vector into task-conditioned training rewards. Extensions include paired clean/fault interventions, marginal fault-discovery rewards, uncertainty-aware evaluation scheduling, and a quarantined cross-domain learning loop. Evaluation emphasizes independently validated fault detection, false alarms, generalization across design families, and total cost to reach a specified quality level. We describe a first experiment on a small, open-tool-compatible benchmark with lightweight models, with full UVM capability admitted through feature-specific qualification. This is a position paper: we specify the framework and the experiments that would test it, and we report no training, EDA, or silicon results.
\end{abstract}

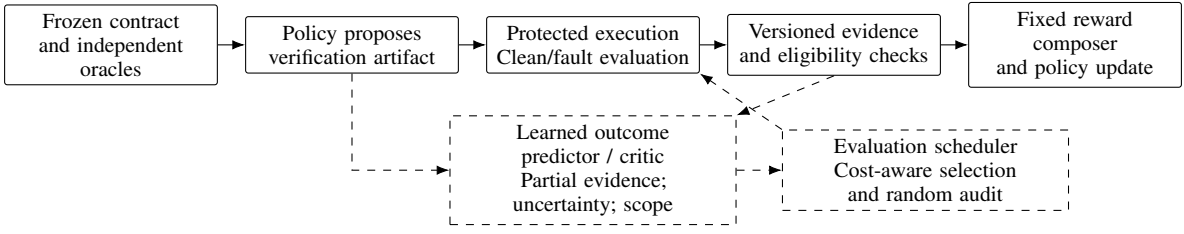
\begin{figure*}[t]
\centering
\begin{tikzpicture}[node distance=0.36cm,>=Latex,
 box/.style={draw,rounded corners=1pt,align=center,minimum height=0.65cm,text width=2.58cm,font=\footnotesize},
 aux/.style={draw,dashed,align=center,minimum height=0.64cm,text width=3.55cm,font=\footnotesize}]
\node[box] (contract) {Frozen contract\\and independent oracles};
\node[box,right=of contract] (policy) {Policy proposes\\verification artifact};
\node[box,right=of policy] (tool) {Protected execution\\Clean/fault evaluation};
\node[box,right=of tool] (evidence) {Versioned evidence\\and eligibility checks};
\node[box,right=of evidence] (reward) {Fixed reward composer\\and policy update};
\draw[->] (contract)--(policy);
\draw[->] (policy)--(tool);
\draw[->] (tool)--(evidence);
\draw[->] (evidence)--(reward);
\node[aux,below=0.55cm of tool] (predictor) {Learned outcome predictor / critic\\Partial evidence; uncertainty; scope};
\node[aux,right=0.60cm of predictor] (scheduler) {Evaluation scheduler\\Cost-aware selection and random audit};
\draw[->,dashed] (policy.south)|-(predictor.west);
\draw[->,dashed] (evidence.south)--(predictor.north east);
\draw[->,dashed] (predictor)--(scheduler);
\draw[->,dashed] (scheduler.north west)--(tool.south east);
\end{tikzpicture}
\caption{Proposed separation of trusted evidence, learned prediction, and budget allocation. Solid paths carry measured evidence to a deterministic reward composer. Dashed paths support selection and critique; they do not replace the contract's required acceptance checks.}
\label{fig:system}
\end{figure*}

\section{Introduction}\label{sec:intro}

Language models can now produce syntactically plausible SystemVerilog, assertions, and testbench components \cite{verilogeval,rtllm,assertllm,autobench}. Turning that ability into a dependable verification assistant requires a training signal that rewards what verification engineers actually value: detection of real design faults, absence of false alarms on correct designs, requirements traceability, and bounded evaluation cost. Simple proxies do not supply that signal. Code coverage can rise while checking weakens; an assertion with zero failures may never have been activated; a testbench that prints an unconditional error ``detects'' every fault and every clean design alike. A policy optimized against such proxies will exploit them \cite{skalse2022,gao2022}.

We take the reward function for a verification-generation policy to be a mapping \((x,y,e)\mapsto R\), where \(x\) describes a verification problem, \(y\) is a generated verification artifact, and \(e\) contains evidence from electronic design automation (EDA) tools. Unlike conversational preference learning \cite{ouyang2022}, this setting has executable ground truth: compilers, simulators, formal engines, and mutation campaigns can measure much of what the reward should encode. But those measurements are expensive, partial, and easy to corrupt. The design problem is therefore not only what to reward but which quantities may be learned, which must remain deterministic, and how partial evidence should be used before a full campaign completes.

This position paper proposes a Verification Reward Model (VRM) framework built on that separation. Fig.~\ref{fig:system} summarizes the architecture. A frozen verification contract and a protected evaluator produce versioned evidence. A learned outcome model predicts expensive future evidence from the contract, the artifact, and explicitly masked partial evidence; a semantic critic identifies evidence-supported weaknesses; a deterministic reward composer maps measured outcomes to training rewards; a cost-aware scheduler decides which evaluations to run. Learned components inform selection and diagnosis. They never override a measured result or an acceptance check.

Our contributions are as follows.
\begin{enumerate}
\tightlist
\item A verification contract and evidence-record formalism that makes observation boundaries, provenance, and evidence status explicit (Sections~\ref{sec:contract} and~\ref{sec:evidence}).
\item A sound fault-detection target for hardware, with validated mutation populations, paired clean/fault controls, attribution rules, and explicit units for recall and false alarms (Section~\ref{sec:fault}).
\item An eligibility-gated reward design with marginal fault-discovery credit and correctly applied potential-based shaping (Section~\ref{sec:reward}).
\item A partial-evidence outcome model, an evidence-linked critic, and an uncertainty-aware evaluation scheduler with mandatory random audits (Sections~\ref{sec:model} to~\ref{sec:uncertainty}).
\item A preregistrable experimental program with hypotheses, family-level splits, baselines, and ablations that could confirm or refute the approach (Section~\ref{sec:experiments}).
\end{enumerate}

We report no training, EDA, or silicon results. The named mechanisms are design labels rather than established techniques, and the numerical thresholds are initial engineering targets to be fixed by preregistration after a pilot.

\subsection{Primary objective}\label{primary-objective}

For a fixed verification task distribution and evaluation budget, we aim to increase the number and diversity of independently validated DUT faults detected while controlling false alarms on independently validated correct behavior. Secondary goals are lower evaluation cost, faster repair, greater requirements traceability, and useful generalization to unfamiliar designs.

The target artifact may be a sequence, test, assertion set, coverage model, scoreboard, monitor, driver, reference model, testplan, or complete environment. Each requires its own evaluation contract. A testplan cannot earn a compilation reward; a stimulus-only sequence cannot independently demonstrate checker correctness.

\subsection{Scope boundaries}\label{scope-boundaries}

The first campaign targets digital block-level verification. Analog behavior, electrical PHY correctness, production signoff, and complete SoC correctness remain outside the initial claim. A selected cycle-level simulator can evaluate a particular digital model; it cannot establish all physical behavior. Every result states its observation boundary and modeling assumptions.

\section{Related Work}\label{sec:related}

\textbf{Preference-based reward models.} InstructGPT trained a reward model from ranked responses and optimized a policy against it \cite{ouyang2022}. DPO removes the separately trained reward model from the policy-optimization loop \cite{dpo2023}, and GRPO provides a group-relative policy-gradient baseline suited to verifiable rewards \cite{grpo2024}. Optimizing against an imperfect reward model degrades true performance as optimization pressure increases \cite{gao2022}, and reward hacking has been characterized formally \cite{skalse2022}. Reward-model ensembles mitigate but do not eliminate the problem \cite{ensembles2023}. LLM-as-a-judge evaluation agrees well with human preference on conversational tasks \cite{llmjudge2023}, but that agreement is measured on tasks without executable ground truth. Our design responds to these results by keeping measured acceptance outside the learned model and by treating optimization against the surrogate as an experiment to be audited rather than a default.

\textbf{LLMs for hardware design and verification.} VerilogEval and RTLLM benchmark RTL generation with fixed functional testbenches \cite{verilogeval,rtllm}. AssertLLM generates SystemVerilog assertions from specification documents \cite{assertllm}, AutoSVA generates formal properties for module interactions from annotated interfaces \cite{autosva}, and AutoBench generates and evaluates HDL testbenches with LLMs \cite{autobench}. These works measure artifact quality by fixed test outcomes or by agreement with reference assertions. None trains a reward model over EDA evidence, and none addresses the credit-assignment problem that arises when the artifact under evaluation is itself the observer.

\textbf{Mutation testing.} Mutation analysis is a mature technique for measuring test-suite adequacy \cite{jia2011}. MCY applies it to hardware with formal filtering of equivalent mutants \cite{mcy}. MIST-RL uses incremental mutation-kill rewards to train unit-test generators for software \cite{mist2026}. We extend the mutation reward to hardware, where valid faults must be qualified under contract conditions, detection must be attributed through temporal checkers, and clean controls must guard against unconditional failure.

\textbf{Reward shaping and uncertainty.} Potential-based shaping preserves optimal policies under stated assumptions \cite{shaping1999}. Conformal prediction supplies distribution-free intervals under exchangeability \cite{conformal2021}. We use both with their assumptions stated explicitly, because adaptive policy optimization violates them in ways that matter for this application.

\section{Contract, Evidence, and Decision Roles}\label{sec:contract}

\subsection{Verification contract}\label{verification-contract}

We define a task contract

\fitdisplay{
C=(S,D,I,O,A,G,B,V),
}

where \(S\) is the requirement set, \(D\) identifies DUT versions, \(I\) is the legal input space, \(O\) defines observability, \(A\) contains environment assumptions, \(G\) identifies trusted oracles and goals, \(B\) is the evaluation budget, and \(V\) records versions and provenance. Interface timing, reset semantics, supported parameter values, allowed nondeterminism, and outstanding-transaction rules belong in \(C\).

The policy may edit only declared artifact paths. It cannot change the DUT, reference model, grading scripts, required coverage bins, mutation population, pass criteria, or tool constraints unless the task explicitly concerns that object and supplies an independent evaluator. Contract changes create new task versions.

\subsection{Five distinct services}\label{five-distinct-services}

Fig.~\ref{fig:system} separates five services. We refer to the two learned components, the outcome model and the semantic critic, jointly as the VRM.

\textbf{Trusted evaluator \(T\).} Executes versioned tools and produces evidence \(e=T(C,y)\). It owns process exit codes, time limits, trace collection, and protected reference files.

\textbf{Outcome model \(O_\theta\).} Predicts future measurements under a named evaluation protocol. At evidence stage \(k\):

\fitdisplay{
p_\theta(q_{>k}\mid C,y,e_{\le k},m_k).
}

The mask \(m_k\) identifies what has actually been observed. The model must never receive its target measurements through logs, filenames, summaries, or metadata. Its scalar utility head, written \(u_\theta\), is used for ranking and scheduling.

\textbf{Semantic critic \(J_\phi\).} Produces a structured assessment of requirements alignment, artifact architecture, and likely gaps. Each assertion is linked to a requirement, code location, or evidence record. Unverified diagnoses are labeled hypotheses.

\textbf{Reward composer \(U\).} Applies fixed validity rules and a versioned scalarization to measured outcomes. It can expose a separate predicted utility for scheduling. A predicted utility cannot overwrite a measured result.

\textbf{Evaluation scheduler \(A_\psi\).} Chooses the next candidate, test subset, or fidelity level to spend budget on. Its objective is evidence acquisition and decision quality. It has no authority to change acceptance criteria.

These services may eventually share representations, but their permissions, training targets, and validation sets remain distinguishable.

\subsection{Why a learned model is useful}\label{why-a-learned-model-is-useful}

A network adds little value when every input already contains all the numbers needed by a fixed reward formula. Its useful jobs are predicting unrun regressions, interpreting local code/evidence interactions, ranking candidates under partial observations, estimating repairability, and deciding which additional measurement is informative. The deterministic composer remains the baseline to beat.

\section{Evidence Representation and Trust Boundaries}\label{sec:evidence}

An evidence record stores a contract digest, candidate digest, toolchain digest, evaluation stage, observed timestamps, workload or seed identifier, measurement units, status, uncertainty, and raw artifact references. Records are append-only. Candidate-controlled text is treated as data, including comments and logs that contain instructions to a judge.

We use distinct statuses: \textbf{pass}, \textbf{fail}, \textbf{unknown}, \textbf{not\_run}, and \textbf{infrastructure\_error}. A formal timeout normally produces unknown. A crashed worker produces an infrastructure error. A reproducible candidate-induced deadlock may be a task failure if the contract defines a completion bound. The outcome depends on its cause and contractual meaning.

\subsection{Structured evidence channels}\label{structured-evidence-channels}

Compilation captures errors, supported language features, elaboration status, and top-level binding. Simulation captures completion reason, transactions driven and observed, checker activation, mismatches, dropped responses, reset episodes, and seed identity. Coverage includes raw numerator, denominator, exclusions, hierarchy, and coverage-model version. Assertion evidence includes attempts, nonvacuous activations where measurable, failures, reset-disable intervals, and pending obligations. Mutation evidence includes validation status, fault family, detection mechanism, and reproducer. Runtime evidence separates queueing, compilation, simulation, parsing, and model inference.

An assertion with zero failures can be inactive. A scoreboard with zero mismatches can have compared zero transactions. A coverage percentage can increase after its denominator is weakened. These conditions must be visible to the evaluator.

\subsection{Evidence stages}\label{evidence-stages}

Stage E0 contains only the contract and generated artifact. E1 adds lint and compilation. E2 adds a short, fixed smoke suite. E3 adds a larger clean regression and requirements coverage. E4 adds validated fault campaigns. E5 adds an independently administered audit on hidden fault families, alternate clean implementations, or a second qualified tool flow.

We train separate heads or stage-conditioned models for E0-to-E4, E1-to-E4, and E2-to-E4 prediction, and include real partial trajectories in training. Random evidence dropout can regularize a model, but it does not by itself replicate the missingness patterns of a scheduler that selectively runs expensive jobs.

\subsection{Repository-scale context}\label{repository-scale-context}

We use bounded retrieval around interfaces, transaction definitions, the candidate diff, requirements, checker call sites, and relevant dependency slices, and record which files were omitted. A flat concatenation of an entire RTL repository can hide truncation-induced failure. We begin with features and compact code encodings, then test whether richer retrieval or a graph encoder improves held-out decisions.

\section{Fault Detection as a Qualified Target}\label{sec:fault}

\subsection{Validated fault population}\label{validated-fault-population}

Let \(M_C\) be the fault population declared for contract \(C\). A mutation is eligible only after checking that it builds, is relevant to legal operating conditions, and changes an observable requirement under those conditions. Where feasible, formal comparison or a witness trace validates relevance. Equivalent, unreachable-under-contract, invalid, and unresolved mutations are separate categories. A timeout is not proof that a mutation is equivalent. MCY provides an implementation starting point for hardware mutation generation and formal filtering \cite{mcy}.

Mutation operators should span state transitions, counter boundaries, handshake retention, reset behavior, address decoding, ordering, data corruption, arbitration, and parameter edge cases. Structural netlist mutations and semantic RTL fault injections are complementary campaigns. Real historical defects with independently reviewed fixes form an additional, more deployment-relevant test set when authorized data exists.

\subsection{Sound kill definition}\label{sound-kill-definition}

A fault is counted as detected when a candidate verification artifact produces a reproducible, contract-relevant failure on a validated faulty DUT and behaves acceptably on paired clean controls. A missing library or simulator crash is not a valid detection. A candidate that prints an unconditional error will also fail the clean control and loses eligibility.

For a test \(t\) and fault \(m\), we define \(K(t,m)=1\) only after detection attribution. The reproducer stores the triggering stimulus, observation, expected behavior, checker identity, and minimized evidence slice. Duplicate manifestations of the same root fault do not become multiple discoveries.

\subsection{Recall and false alarms use explicit units}\label{recall-and-false-alarms-use-explicit-units}

For a fixed set of independently adjudicated design/workload episodes, define a positive episode as one containing a qualifying defect. Precision and recall then have their usual meanings:

\fitdisplay{
P=\frac{TP}{TP+FP},\qquad R=\frac{TP}{TP+FN}.
}

The same episode definition must govern all four counts. Combining a count of unique buggy designs with a count of millions of clean transactions does not yield a precision. We report mutation recall by fault family and clean false-alarm rate separately. F1 may be a secondary diagnostic; its value depends on the artificial defect prevalence in the benchmark.

Zero observed clean failures does not establish a zero false-alarm probability. Under an independent Bernoulli model, zero failures in \(n\) trials gives a one-sided 95\% upper bound \(1-0.05^{1/n}\). For \(n=100\), this is about 2.95\%. Correlated seeds require cluster-aware analysis and can supply much less information than 100 independent trials.

\section{Reward Design}\label{sec:reward}

\subsection{Quality vector and ranges}\label{quality-vector-and-ranges}

The measured quality vector is

\fitdisplay{
q=(q_{\mathrm{fault}},q_{\mathrm{req}},q_{\mathrm{robust}},q_{\mathrm{eff}}),
\qquad q_j\in[0,1].
}

Here fault quality is severity-weighted detection over a fixed validated population; requirements quality measures independently instrumented, checked obligations; robustness measures predefined legal stress scenarios; and efficiency is a bounded transformation of runtime at an otherwise comparable quality level. Raw code coverage and subjective style remain diagnostics in the initial experiment.

Weights satisfy \(w_j\ge0\) and \(\sum_jw_j=1\). A starting scalarization for complete environments is \((0.55,0.25,0.15,0.05)\). These values are hypotheses, not empirically optimal settings. Scoreboard, assertion, stimulus, and testplan training each use a different versioned contract. A weight is interpretable only together with the component's normalization and denominator.

\subsection{Eligibility before utility}\label{eligibility-before-utility}

We define the measured terminal reward as

\fitdisplay{
R_{\mathrm{DV}}=
\begin{cases}
-1,&\text{tampering or confirmed unsoundness},\\
-0.5,&\text{attributable build/task failure},\\
\mathrm{NA},&\text{required evidence unresolved},\\
\sum_jw_jq_j,&\text{eligible evidence complete}.
\end{cases}
}

The numerical range of scored records is \([-1,1]\). NA means the sample is masked from the applicable outcome loss or held for another evaluation; it is not silently converted to zero. The controller budgets unresolved work so that deliberate resource exhaustion cannot create an advantageous free pass.

Eligibility includes protected-file integrity, correct binding, minimum activity, clean-control behavior, reference-oracle validity, and all task-required evidence. Failure on a known faulty DUT is expected evidence of success when attribution is valid. A global rule that penalizes every simulation failure is therefore unsuitable for verification generation.

\subsection{Marginal suite utility}\label{marginal-suite-utility}

For a suite \(S\) and fixed nonnegative fault weights \(v_m\) with \(\sum_m v_m>0\), define

\fitdisplay{
F(S)=\frac{\sum_{m\in M_C}v_m\,\mathbf{1}[\exists t\in S:K(t,m)=1]}{\sum_{m\in M_C}v_m}.
}

The incremental value of a new test is \(F(S\cup\{t\})-F(S)\). Repeated tests that detect the same already-covered faults receive no additional discovery credit. All denominators are frozen for a campaign. This principle has precedent in software mutation-based RL \cite{mist2026}; the hardware extension must establish valid faults, temporal behavior, and checker soundness.

Runtime cost can be included as a separately bounded penalty or a budget constraint. We report both unconstrained detection and cost-normalized detection so that smaller tests do not win by omitting required behavior.

\subsection{Potential-based shaping}\label{potential-based-shaping}

Intermediate feedback can use a fixed potential over observable artifact states:

\fitdisplay{
r'_t=r_t+\gamma\Phi(s_{t+1})-\Phi(s_t).
}

Potential-based shaping has policy-invariance results under the associated MDP assumptions \cite{shaping1999}. For the finite repair episode considered here, we freeze \(\Phi\) during each collection/update cycle and set the terminal potential consistently, normally to zero. Clipping individual shaping terms, changing the potential mid-episode, or rewarding only positive improvements can invalidate the intended telescoping argument. Final acceptance uses the original terminal metric.

\section{Model Architecture and Learning Objectives}\label{sec:model}

\subsection{Baselines first}\label{baselines-first}

We implement a fixed-rule composer, a tabular outcome predictor, and a frozen generic LLM judge before a domain-trained VRM. The tabular predictor sees only information genuinely available at its evaluation stage. A gradient-boosted tree or small multilayer perceptron may outperform an expensive text model on a small dataset. That outcome would itself be informative and would argue for the simpler model.

\subsection{Structured outcome model}\label{structured-outcome-model}

A candidate architecture combines a compact code/specification encoder, structured EDA features, evidence masks, and task/fidelity embeddings. Late fusion feeds classification, regression, ranking, and abstention heads. The first version need not use a graph neural network. A later requirement-to-code-to-checker graph can expose whether stimulus, observation, and comparison paths jointly support an obligation.

We predict per-fault detection probabilities where the data supports them and aggregate by fault family with uncertainty. For count outcomes, successes and trials are retained instead of collapsed into a percentage. A beta-binomial head is one candidate for overdispersed detection counts. Its assumptions must be checked against simpler calibrated models.

\subsection{Multi-task training}\label{multi-task-training}

We use a masked objective

\fitdisplay{
\mathcal{L}=\lambda_b\mathcal{L}_{\mathrm{BCE}}
+\lambda_r\mathcal{L}_{\mathrm{reg}}
+\lambda_p\mathcal{L}_{\mathrm{pair}}
+\lambda_c\mathcal{L}_{\mathrm{critic}}.
}

Binary heads estimate declared events, continuous heads predict future metrics, pairwise heads rank candidates, and the optional critic head learns evidence-linked feedback. Loss weights are tuned on development families. Expert labels are recorded with reviewer disagreement, not collapsed into an unexplained authoritative score.

For a valid pair \(a\succ b\), the ranking term uses the utility head \(u_\theta\) of the outcome model:

\fitdisplay{
\mathcal{L}_{\mathrm{pair}}=-\log\sigma(u_\theta(a)-u_\theta(b)).
}

Pairs must share contract, fidelity, budget, and scalarization. Pareto-incomparable candidates remain incomparable unless a stakeholder weight vector is supplied. A higher-mutation/lower-coverage candidate is not automatically ranked above every alternative. Near ties and unresolved evidence are masked or modeled explicitly. Many pairs from the same candidates do not create independent data.

\subsection{Critic reliability}\label{critic-reliability}

The critic emits issue category, requirement identifier, code span, evidence identifier, confidence, and a proposed discriminating test. An unsupported explanation cannot count as a verified defect. We evaluate whether following a critique improves independently measured results. A fluent explanation with no repair benefit should not increase reward.

\section{Counterfactual and Interaction-Based Extensions}\label{sec:counterfactual}

\subsection{Paired clean/fault interventions}\label{paired-cleanfault-interventions}

Matched runs use the same legal stimulus, seed, tool version, and environment while changing only the DUT fault. This isolates the observed effect of that injected change within the controlled experiment. It does not establish general real-world causal attribution beyond the model and contract.

A second intervention changes only the checker or stimulus component. For example, compare an original sequence and an added backpressure sequence under a weak and a repaired scoreboard. A four-cell experiment reveals whether the stimulus improvement becomes observable only after the checker is repaired. This addresses a central credit-assignment problem: a useful test can appear worthless when the observer cannot detect its consequence.

\subsection{Requirement-fault-checker interaction record}\label{requirement-fault-checker-interaction-record}

We store a sparse tensor indexed by requirement, fault family, stimulus family, checker, and evidence stage. Its entries contain activation and detection outcomes. The VRM is trained to distinguish failure to activate a condition, failure to propagate a fault, and failure to observe or check the resulting behavior. These are explanatory targets; supervision must come from instrumentation or reviewed traces.

\subsection{Metamorphic soundness tests}\label{metamorphic-soundness-tests}

We challenge the artifact with semantics-preserving renaming, legal timing variation, alternate clean implementations, harmless logging changes, and reordered independent transactions where the contract permits it. A black-box verification task should not rely on private internal signal names. A white-box task can use declared internal observability and must identify that dependency.

These challenges also test the judge: harmless formatting should not materially change its correctness estimate; removing a real comparison should lower it. An invariance failure becomes training data and an evaluator defect ticket.

\section{Uncertainty-Aware Evaluation and Audit}\label{sec:uncertainty}

\subsection{What uncertainty means}\label{what-uncertainty-means}

We separate aleatoric variation, such as random workload behavior, from epistemic uncertainty caused by unfamiliar artifacts or sparse data. Ensembles, quantile models, and held-out calibration are candidate techniques. Ensemble agreement alone cannot establish correctness; reward-model ensembles are known to retain reward-hacking risk \cite{ensembles2023}.

Conformal intervals can be considered when calibration and deployment data satisfy the relevant exchangeability assumptions \cite{conformal2021}. Adaptive policy optimization changes the distribution. Ordinary calibration guarantees cannot be asserted automatically for newly optimized candidates or unseen IP types.

\subsection{Scheduler objective}\label{scheduler-objective}

The scheduler selects an evaluation action \(a\) using an estimated decision-value objective such as

\fitdisplay{
A(a)=\frac{\mathbb{E}[\text{reduction in selection regret}\mid a]}{\mathbb{E}[\text{evaluation cost}\mid a]}.
}

A practical approximation combines expected quality improvement, uncertainty, and fault-family diversity. The approximation is a scheduling heuristic. It is not a theorem that the next selected candidate is optimal.

We reserve an initial 20\% of evaluation budget for stratified random audits, independent of predicted quality, and record selection probabilities. This limits blind spots and makes selection bias measurable. Predictive screening is allowed to save EDA work only after its false-rejection behavior is audited. All release candidates receive the mandated full evaluation.

\subsection{Stopping and escalation}\label{stopping-and-escalation}

Evaluation escalates to real measurement when uncertainty is high, evidence conflicts, a new failure family appears, or a candidate approaches a release threshold. Evaluation of a candidate stops when it is conclusively invalid or cannot meet the remaining budget. Infrastructure failures are retried under a bounded policy and counted in operational cost.

\section{Policy Training and the Data Flywheel}\label{sec:policy}

We begin with supervised fine-tuning on authorized, externally accepted artifacts and repair trajectories. DPO is a useful offline preference baseline that does not require a separately optimized reward model during policy training \cite{dpo2023}. For online RL, we evaluate a GRPO-style or comparable policy-gradient baseline using fixed external rewards \cite{grpo2024}. Algorithm choice is an empirical question settled on the development families.

We store the exact behavior-policy checkpoint, prompt, sampling settings, actions, tool outputs, and reward version. For long EDA jobs, a complete small batch is gathered under a frozen policy and scorer before updating. A GRPO trainer cannot treat arbitrary stale trajectories as on-policy without appropriate handling. Groups with identical rewards carry no useful relative preference signal; the remedy is harder tasks or more diverse data, not invented differences.

The initial role of the VRM is reranking and scheduling. Its use as a policy-optimization reward is a later ablation with a small, audited influence. Measurements and predictions remain separate fields. Repeated optimization against a surrogate is specifically tested for divergence between predicted utility and hidden external outcomes, the failure mode documented by Gao et al. \cite{gao2022}.

\subsection{Flywheel sequence}\label{flywheel-sequence}

\begin{enumerate}
\tightlist
\item
  Freeze a contract set, evaluator, split, and policy checkpoint.
\item
  Generate multiple bounded candidates or repair actions.
\item
  Run integrity, build, and activity checks.
\item
  Allocate clean, stress, and fault evaluations using the scheduler plus random audits.
\item
  Adjudicate ambiguous failures and compute measured rewards.
\item
  Train the outcome model and critic on training families only.
\item
  Train a challenger policy using the selected baseline or RL procedure.
\item
  Evaluate the challenger on development families with frozen external metrics.
\item
  Promote only after a separate release audit; otherwise retain the incumbent.
\item
  Add genuinely new, authorized failures to the next training cycle and version all changes.
\end{enumerate}

The hidden final test suite is not fed back into the same development cycle. Once detailed test feedback has influenced development, that suite is retired from final-test status.

\section{Worked Verification Scenario}\label{sec:scenario}

Consider a parameterized synchronous FIFO whose contract fixes reset behavior, simultaneous push/pop semantics, ordering, and backpressure. Candidate A drives many random transactions and reaches broad RTL activity, but compares only a total transfer count. Candidate B exercises wraparound and compares every returned data item using an independent queue model.

A validated fault duplicates one stored value at wraparound. Candidate A may retain a high activity score and miss the defect. Candidate B produces a data mismatch on the faulty FIFO and passes paired clean runs. Its mismatch is positive verification evidence. A third candidate emits an error unconditionally; its clean-control failures disqualify it.

The VRM should predict the relative detection strength from the comparison logic and smoke evidence before a full fault campaign. Its critic can identify the missing data comparison in A, cite the affected requirement, and propose a wraparound reproducer. The final detection credit is awarded only after the trusted campaign validates that behavior.

\section{Experimental Program}\label{sec:experiments}

\subsection{Hypotheses}\label{hypotheses}

H1: partial-evidence outcome prediction improves candidate selection under a fixed total evaluation budget. H2: soundness-qualified mutation targets generalize to withheld fault operators and historical defects better than raw coverage targets. H3: component-intervention supervision improves repair credit assignment. H4: learned critiques improve externally measured repairs beyond a generic judge. H5: the complete flywheel improves a frozen-policy baseline without increasing clean false alarms.

\subsection{Splits and baselines}\label{splits-and-baselines}

We group all descendants of a design, generator template, reference implementation, and mutation family before splitting, and maintain train, development, calibration, and sealed-test partitions. Evaluation covers both new implementations of familiar IP types and withheld IP categories. Public benchmark scores are supplementary because exposure during model pretraining may be unknown.

Baselines are random candidate selection, fixed heuristics, generic LLM judging, tabular prediction, text-only VRM, partial-evidence VRM, and a full-evidence oracle-ranking upper reference. The oracle consumes the expensive measurements and must be charged their actual cost when used operationally.

\subsection{Metrics}\label{metrics}

Primary metrics are validated fault recall at a bounded false-alarm rate and total cost to reach a target recall. We report per-family recall, top-\(k\) selection regret, missed-promising-candidate rate, calibration, coverage of prediction intervals, abstention rate, repair success, and valid-output rate. Cost accounting includes model inference, data labeling, retraining, tool startup, and failed jobs.

We use paired comparisons on the same held-out contracts and bootstrap at the design-family level, reporting effect sizes and confidence intervals. A high Pearson or Spearman correlation alone does not demonstrate safe screening. The consequences of the actual top-\(k\) decisions are what is evaluated.

\subsection{Critical ablations}\label{critical-ablations}

Remove mutation validity filtering; remove clean controls; replace requirement coverage with raw coverage; remove evidence masks; compare family-level and random-record splits; remove random audits; remove the critic; replace the text model with tabular features; optimize aggressively against the surrogate. The first several ablations are intentionally unsafe diagnostic experiments and cannot become deployment configurations.

\section{Resource Model and First Implementation}\label{sec:resources}

The first campaign uses CPU EDA workers and a small local model before any large-model campaign. A single consumer-grade GPU can support an optional compact-model path; fit and throughput must be measured on the target hardware. Full-model or large-context training is not assumed. QLoRA provides a parameter-efficient training technique; actual fit and throughput still require local measurement \cite{qlora2023}.

For \(D\) designs, \(N\) candidates, \(S\) clean runs, \(M\) faults, and \(S_m\) runs per fault, the simulation count is

\fitdisplay{
N_{\mathrm{sim}}=DN(S+MS_m).
}

An illustrative pilot with \(D=24\), \(N=6\), \(S=3\), \(M=24\), and \(S_m=2\) has 7,344 simulation instances. At an assumed five seconds per simulation this is 10.2 CPU-hours of simulation alone. Mutant compilation, original compilation, formal filtering, scheduling, and model work are additional. These are planning calculations, not measured runtimes.

The pilot starts with synthesizable SystemVerilog, a supported assertion subset, and independent testbench infrastructure. Verilator and SymbiYosys supply relevant open-source capabilities \cite{verilator,sby}. UVM support must be qualified for the exact pinned features and library version; the existence of upstream UVM work in Verilator does not justify an assumption that every existing environment will run unchanged \cite{uvmverilator}.

\section{Cross-Domain Extension and Independence}\label{sec:crossdomain}

A verification model can evaluate artifacts used to test generated RTL. The RTL generator and verification generator must not jointly redefine their own oracle. Trusted functional behavior is frozen and reference models are independently developed or reviewed. Newly discovered RTL counterexamples enter a quarantine set until adjudicated. Architecture requirements can supply obligations to both tracks, but shared generated specifications are not independent proof of intent.

Common contract and evidence schemas are shared across tracks while acceptance owners remain separate. This enables shared data engineering and representation learning without merging correctness authority into a single self-evaluating model.

\section{Failure Modes and Research Decisions}\label{sec:failures}

The largest risks are weak oracles, mutation distribution mismatch, hidden label leakage, evaluator tampering, false failures from incorrect checkers, unsupported language constructs, and uncalibrated surrogate optimization. Each corresponds to an ablation or metamorphic challenge in Sections~\ref{sec:counterfactual} and~\ref{sec:experiments}. A learned model that cannot beat the tabular baseline should be simplified or deferred. A fault reward that improves synthetic mutations while harming historical-defect performance must be reweighted or redesigned.

The strongest initial contribution would be a reproducible evidence protocol plus an audited cost-quality gain on unfamiliar design families. A general verification foundation model is a longer-term outcome that requires broader data and demonstrated transfer. The immediate scientific claim should remain narrower and measurable.

\section{Conclusion}\label{sec:conclusion}

The proposed VRM framework treats verification quality as a contract-dependent, evidence-supported quantity. Learned models predict missing outcomes, interpret artifact weaknesses, and allocate evaluation effort. Trusted external checks determine eligibility and release. The recommended first experiment is a small family-split FIFO, arbiter, register-block, and stream-controller campaign that tests whether partial-evidence ranking reduces EDA cost without losing validated fault detection. That experiment can establish the value of the learning layer before committing to a larger domain-specific policy.

\balance

\end{document}